\pdfoutput=1
\documentclass[conference]{IEEEtran}
\IEEEoverridecommandlockouts
\usepackage{cite}
\usepackage{amsmath,amssymb,amsfonts}
\usepackage{algorithmic}
\usepackage{graphicx}
\usepackage{url}
\usepackage{textcomp}
\usepackage{xcolor}
\usepackage{listings}
\usepackage{xcolor} 
\def\BibTeX{{\rm B\kern-.05em{\sc i\kern-.025em b}\kern-.08em
    T\kern-.1667em\lower.7ex\hbox{E}\kern-.125emX}}
\begin{document}

\title{\LARGE Testing LLMs on Code Generation with \\ Varying Levels of Prompt Specificity}

\author{\IEEEauthorblockN{Lincoln Murr, Morgan Grainger, and David Gao}
\IEEEauthorblockA{\textit{Department of Computer Science},
\textit{Vanderbilt University, Tennessee},
Nashville, TN, USA \\
\{lincoln.d.murr, morgan.j.grainger, david.gao\}@vanderbilt.edu}}

\maketitle

\begin{abstract}
Large language models (LLMs) have demonstrated unparalleled prowess in mimicking human-like text generation and processing. Among the myriad of applications that benefit from LLMs, automated code generation is increasingly promising. The potential to transform natural language prompts into executable code promises a major shift in software development practices and paves the way for significant reductions in manual coding efforts and the likelihood of human-induced errors. 

This paper reports the results of a study that evaluates the performance of various LLMs, such as Bard, ChatGPT-3.5, ChatGPT-4, and Claude-2, in generating Python for coding problems. We focus on how levels of prompt specificity impact the accuracy, time efficiency, and space efficiency of the generated code. A benchmark of 104 coding problems, each with four types of prompts with varying degrees of tests and specificity, was employed to examine these aspects comprehensively. Our results indicate significant variations in performance across different LLMs and prompt types, and its key contribution is to reveal the ideal prompting strategy for creating accurate Python functions. This study lays the groundwork for further research in LLM capabilities and suggests practical implications for utilizing LLMs in automated code generation tasks and test-driven development.
\end{abstract}


\section{Introduction}
\textbf{Emerging trends and challenges.}
The advent of large language models (LLMs) are transforming  computational science and technology, offering unprecedented capabilities in processing and generating human-like text. LLMs hold promise in applications ranging from natural language understanding to automated code generation. Automated code generation is a field that explores the potential of LLMs to convert natural language prompts into executable computer code, potentially transforming software development by significantly reducing manual coding effort and minimizing human error. For example, 40\% of the code generated by GitHub Copilot is being checked in with no modifications ~\cite{Microsoft_2023}.  Evaluating the effectiveness and versatility of LLMs in code generation tasks is complicated, however, necessitating meticulous exploration and comprehensive testing methods.

The evolution of generative AI has been marked by the release of LLMs, such as Bard, ChatGPT-3.5, ChatGPT-4, and Claude-2, each with unique architectures and capabilities. These LLMs are driving innovations in natural language processing (NLP) and machine learning research in diverse domains. Applying these LLMs to generate code automatically is enabling more intuitive and efficient software development processes, allowing developers to focus on higher-level designs and logic. However, the inherent complexity and variability of natural language pose significant challenges in harnessing the full potential of LLMs in this domain.

\textbf{Overview of prompt engineering.}
A key element in leveraging LLMs for code generation is \textit{prompt 
engineering}~\cite{promptengineering}, which involves crafting prompts with varying degrees of specificity to elicit desired responses from LLMs. A prompt is typically a statement, question, or instruction given to an LLM to invoke a specific response or action. It guides the LLM in understanding the user’s requirements and generating coherent and contextually appropriate responses. For instance, in code generation tasks, a prompt could range from a high-level problem description to a detailed input, output, and functional requirements specification. Understanding the intricacies of prompts and prompt engineering is crucial to enabling the customization of user interactions with LLMs, allowing them to optimize results in accordance with the requirements of a given task.

Prompt specificity can significantly impact the accuracy and efficiency of the  output generated by an LLM ~\cite{si2023prompting}. This specificity refers to the level of detail and clarity provided in the information given to an LLM. A high-specificity prompt may include explicit instructions, detailed requirements, and clear expectations, leaving little room for interpretation or ambiguity. Conversely, a low-specificity prompt is more general, open-ended, and leaves more to the interpretation by the LLM.

The level of specificity in a prompt can significantly impact the output generated by LLMs. A highly specific prompt that provides an LLM with unambiguous directions may yield outputs that are precise, accurate, and closely aligned with user expectations. However, it may also constrain an LLM's creative and generative capabilities, potentially leading to overly restrictive outputs and a lack of innovation or adaptability.

Conversely, a prompt with lower specificity allows an LLM more freedom to interpret user intentions and generate more informative responses. This flexibility can result in more creative, diverse, and adaptable outputs in different contexts. However, it also poses the risk of an LLM generating off-target, irrelevant, or incorrect outputs due to the lack of clear guidance and constraints.

\textbf{Overview of test-driven development}
Test-driven development (TDD)~\cite{TDD} is a widely adopted practice in software engineering where developers write tests before writing the actual code. It is based on developing and refining software by first writing tests to define the desired behavior of the code and then writing code to pass those tests. Integrating TDD principles in prompt engineering can be particularly helpful in shaping LLM responses. 

When LLMs are provided with tests as part of the prompts, they can serve as a definitive specification of what the code should accomplish, guiding the model to generate code that meets the stipulated requirements and passes the provided tests. Incorporating these tests adds a layer of specificity and constraint to the prompts, potentially aiding in generating more precise and robust code while also offering insights into LLMs' adaptability and comprehension skills in conforming to established software development practices.  Test-enabled prompts can significantly impact the accuracy, efficiency, and reliability of the code generated by LLMs, ensuring that the generated code is syntactically correct, functionally sound, and adherent to the specified behavior.  

\textbf{Summary of research objectives, questions, and contributions.}
The main objective of our research is to determine the effectiveness of including tests in prompts to LLMs to guide their generation of code.  In particular, we  assess the performance of popular LLMs, such as ChatGPT-3.5, ChatGPT-4, Bard, and Claude-2, on a diverse set of coding problems, each presented with varying levels of prompt specificity. A secondary objective of this study is to explore the performance of each LLM on these coding problems to evaluate their capabilities.

Central to this study are the following research questions:
\begin{enumerate}
    \item How do varying levels of prompt specificity influence the quality, accuracy, and efficiency of the code generated by LLMs?
    \item Is there a discernible difference in performance across various LLMs in code generation tasks?
    \item Does employing a test-driven prompting approach lead to different---possibly more optimized---solutions compared to traditional prompting techniques?
\end{enumerate}

Our research focuses on exploring how different levels of prompt specificity impact the performance of LLMs in generating Python code. By examining a diverse range of coding problems and employing varying degrees of prompt specificity, this study evaluates the reasoning capabilities and limitations of leading LLMs such as ChatGPT-3.5, ChatGPT-4, Bard, and Claude-2. Understanding how specificity in prompts influences generated code provides insights into these LLMs' adaptability and comprehension skills, revealing their ability to interpret and respond to varying levels of information detail.

A key contribution of this paper is the illumination of the inherent reasoning capabilities of current LLMs. By interacting with these LLMs using prompts of varying specificity and analyzing the resultant code, we aim to discern how these models can comprehend, interpret, and generate logical and functional Python code. This exploration sheds light on the depth of understanding and logical reasoning these LLMs possess, offering a glimpse into their capabilities in code generation and prompt interpretation.

This paper also provides a well-founded position on the optimal manner of prompting LLMs for generating Python code. By evaluating the accuracy, efficiency, and reliability of the code generated under different prompting strategies, we codify which approach yields the most desirable outcomes. This insight is crucial as it informs best practices in leveraging LLMs for automated code generation, allowing for more effective and reliable utilization of these LLMs in practical software development scenarios.

The findings of this study have implications for both practical applications and future research. By determining the most effective prompting strategies, developers and researchers can optimize the use of LLMs in software development, enhancing productivity and reducing the likelihood of errors. Moreover, this research lays a foundation for further exploration into the capabilities of LLMs, spurring further investigations into how these LLMs can be refined and improved to understand better and generate complex logical programming constructs.

\textbf{Paper organization.}
This paper is organized into the following sections:
 \begin{itemize}
     \item introduction (Section I): Provides background on LLMs for code generation and prompt engineering, summarizes the research objectives, questions, and contributions, and gives an overview of the paper structure.
    \item Related Work (Section II): Reviews relevant literature on prompt engineering for LLMs and code generation benchmarks.
    \item Testing Methodology (Section III): Outlines the approach for generating coding problems, crafting varied prompt types, testing the LLMs, and evaluation metrics.
    \item Analysis of Results (Section IV): Presents detailed analysis of test results by prompt type, overall model performance, comparisons between models, and implications of findings.
    \item Future Work (Section V): Discusses promising avenues for future research, including evaluating different LLMs, programming languages, more complex problems, prompt types, and existing benchmarks.
    \item Concluding Remarks (Section VI): Summarizes key findings and contributions, limitations, and opportunities to build on this research.
    \item Appendix: Provides access to supplemental information like test cases and prompts.
 \end{itemize}

\section{Related Work}

This section summarizes recent work on prompt engineering and the use of generative AI models for software engineering.

Applying LLMs for software engineering has garnered increasing attention as these transformer-based models have surged in popularity for general usage. Evaluation of different models spans a range of programming tasks such as Parsons Problems \cite{10.1145/3587102.3588805}, CS1 Problems \cite{denny2022conversing}, Data Science topics \cite{pmlr-v202-lai23b}, and competitive programming problems \cite{doi:10.1126/science.abq1158}, among others.

Much research, akin to that by Li \textit{et al.}, delves into training or fine-tuning new models specifically for programming tasks. Preliminary results suggest that these models might even outperform humans in certain programming tasks \cite{doi:10.1126/science.abq1158}. As models enhance these capabilities, we aspire that general-purpose large language models readily accessible to the masses—like GPT-4, Bard, and Claude-2—can be assessed. This would enable developers to harness their power without the necessity of training bespoke models.

A significant consideration is the application to distinct programming tasks, which consequently alters the evaluation metrics across studies. While several research initiatives have employed existing programming problem evaluation platforms like Codeforces \cite{doi:10.1126/science.abq1158} and Leetcode \cite{sakib2023extending}, this methodology constrains the variety of problems that can be assessed. It potentially narrows our understanding of the authentic coding proficiencies of large language models. Consequently, Lai \textit{et al.} instituted a benchmark that evaluates code generated by these models beyond mere functional correctness \cite{pmlr-v202-lai23b}. This approach, among other benchmarks, is meticulously dissected in the survey by Zan \textit{et al.}, which critiques 17 benchmarks and their limitations \cite{zan-etal-2023-large}.

The domain of prompt engineering has also piqued interest. This discipline optimizes user input structures to generative models to refine the ensuing output. The rapid evolution of this field is attributed to the realization that outputs from large language models are profoundly influenced by their input sequences and can fluctuate based on various prompting techniques \cite{white2023chatgpt}. Specifically, Reeves \textit{et al.} discerned negligible performance variations when altering prompts for programming tasks within the Parson's problem formats \cite{10.1145/3587102.3588805}. In contrast, a separate study by Denny, Kumar, and Giacaman, which primarily fixated on the prompt engineering dimension, revealed that prompt categorizations yielded varied results yet proffered a novel perspective into the potential diversities in prompting \cite{denny2022conversing}.

In our pursuit, we focus on the ramifications of prompt specificity by adopting a more systematic categorization approach. Moreover, we intend to assess code generation capabilities across a spectrum of publicly accessible, general-purpose large language models. Through our efforts, we aspire to deliver a tangible application that benefits a wider audience, building upon the rich tapestry of preceding research endeavors.

\section{Testing Methodology}

This section provides a detailed overview of the testing methodology, including the process for generating coding problems, crafting varied prompt types to assess LLMs under different conditions, the approach to testing and evaluating the LLMs, and the metrics used to quantify performance.

\subsection{Generation of Coding Problems}

A standardized approach was employed to conduct a meticulous and comprehensive examination of the adaptability and comprehension skills of leading LLMs—GPT-4, Claude-2, and Bard. Each model was presented with the same prompt, included in the appendix, tasking them to generate coding problems. The objective was to understand how well these models could conceptualize and formulate coding problems when provided with identical instructions. Twenty-six problems, each with four prompts, were created - 11 from GPT-4, 10 from Claude-2, and five from Bard. This variety, with the proportion loosely based on the popularity of each model, captures a broader scope of coding questions than one model and provides some protection against an LLM being biased toward its own problems.

\subsection{Creation of Varied Prompts}

In addition to creating coding problems, each model was also instructed to generate four distinct prompts for each problem they created. This was done to present each problem in four unique styles, aiming to assess the adaptability and comprehension skills of the LLMs under varying levels of instruction specificity and clarity. The four styles of prompts created, along with an example of the prompts used for the problem of sorting the even integers in an array, were as follows:

\begin{enumerate}
    \item \textbf{Prompt Only:} This included just the problem statement, providing the minimum amount of information, and evaluated how well the LLMs could interpret and respond to straightforward, unembellished instructions.\\
    \textit{Example:} Implement a function that takes an array of integers and sorts the even numbers in ascending order while leaving the odd numbers in their original positions.

    \item \textbf{Prompt with Tests:} This encompassed a problem statement supplemented with example test cases. This style assessed the LLMs’ ability to interpret and utilize additional contextual information in the form of test cases to generate code.\\
    \textit{Example:} Implement a function that takes an array of integers and sorts the even numbers in ascending order while leaving the odd numbers in their original positions.
    \begin{lstlisting}
Test Cases:
- Input: [5, 3, 2, 8, 1, 4]
  Expected Output: [5, 3, 2, 4, 1, 8]
- Input: [0, 1, 2, 3, 4, 5]
  Expected Output: [0, 1, 2, 3, 4, 5]
    \end{lstlisting}

    \item \textbf{Prompt Tests Only:} In this style, only test cases were provided, albeit including the names of the tests. This tested the LLMs' ability to infer the problem requirements solely based on the test cases and their names, measuring their deductive reasoning skills.\\
    \textit{Example:} Write a Python function to pass the following tests:
    \begin{lstlisting}
    def test_sort_even_numbers():
    assert sort_even_numbers([5, 3, 2, 8, 1, 4]) == [5, 3, 2, 4, 1, 8]
    assert sort_even_numbers([0, 1, 2, 3, 4, 5]) == [0, 1, 2, 3, 4, 5]
    \end{lstlisting}

    \item \textbf{Prompt Generic Tests:} This involved providing test cases with a masked function name, introducing an element of ambiguity, and evaluating the LLMs’ adaptability and ability to generate coherent code despite the absence of explicit function names. This was introduced as a metric to measure an LLM’s advanced reasoning and deduction capabilities and is not expected to have strong results or be replicated for production use.\\
    \textit{Example:} Write a Python function to pass the following tests:
    \begin{lstlisting}
def test_generic_0():
    assert function([5, 3, 2, 8, 1, 4]) == [5, 3, 2, 4, 1, 8]
    assert function([0, 1, 2, 3, 4, 5]) == [0, 1, 2, 3, 4, 5]
    \end{lstlisting}
\end{enumerate}

\subsection{Testing and Evaluation}

Once the prompts and their corresponding problems were generated, the LLMs were subjected to a comprehensive series of tests to gauge their performance and correctness accurately. These tests included additional test cases that were not initially part of the original prompts, ensuring the assessment of extensibility. Each problem was generated in its own LLM session to ensure that prior information did not bias the results or provide more insight into the type of problem the generic testing prompts were evaluating. The tests were designed to measure the number of passes and fails for each prompt type, offering insights into the reliability and accuracy of the generated code under different prompting conditions. The data derived from the tests was categorized based on the prompt and LLM solver, enabling a detailed comparative analysis of the performance of each LLM across different prompt types and problem scenarios.

\section{Analysis of Results}

Section IV analyzes the aggregated test results, examining the performance of the LLMs overall and by model and prompt type. It also draws comparisons between the models and discusses key implications of the findings.

\begin{table*}[t]
    \centering
    \caption{Test Results}
\begin{tabular}{|l|c|c|c|c|c|}
    \hline
    Model & prompt\_generic\_tests.py & prompt\_only.py & prompt\_tests\_only.py & prompt\_with\_tests.py & Overall \\
    \hline
    Bard & 36/63 & 46/63 & 44/63 & 46/63 & 172/252 \\
    Claude & 44/63 & \textbf{54/63} & 50/63 & 56/63 & 204/252 \\
    GPT-3.5 & 48/63 & 52/63 & \textbf{55/63} & 55/63 & \textbf{210/252} \\
    GPT-4 & \textbf{51/63} & 49/63 & 54/63 & \textbf{55/63} & 209/252 \\
    Text-davinci-002 & 21/63 & 21/63 & 28/63 & 16/63 & 86/252 \\
    \hline 
    Total & 200/315 & 222/315 & 231/315 & 228/315 & 881/1260 \\
    \hline
\end{tabular}
\label{tab:results}
\end{table*}

\subsection{Analysis of Overall Model Performance by Prompt Style}

The aggregated results across all models for the various prompt styles offer intriguing insights into the general adaptability and proficiency of LLMs in code generation tasks:

\begin{enumerate}
    \item \textbf{Prompt Tests Only (231/315):} This style emerged as the most effective, garnering the highest total score. The fact that the models performed best when provided solely with clearly defined test cases suggests a strong deductive reasoning capability among LLMs. It implies that the models are adept at inferring the requirements of a problem based purely on the expected outcomes and can generate code that aligns with these inferred requirements. It may also mean that prompting with tests is analogous to the “chain-of-thought” prompt engineering format, in which an LLM is given a set of example cases or questions to refine its reasoning abilities further \cite{wei2023chainofthought}.
    \item \textbf{Prompt with Tests (228/315):} Closely following the "Prompt Tests Only" style, this approach combining problem statements with example test cases also proved highly effective. Combining descriptive instructions and practical examples might offer a balance of guidance and context, enabling LLMs to generate accurate and contextually relevant code. Given the close score to prompting with tests alone, it is difficult to make a conclusive judgment on which format is better, but it might be hypothesized that the prompt with the test could provide conflicting or ambiguous information that is not present when tests alone are inputted.
    \item \textbf{Prompt Only (222/315):} With just the problem statement provided, the models still showcased strong performance, albeit slightly lower than when tests were included. This indicates that while LLMs can effectively interpret and respond to straightforward problem descriptions, the additional context of test cases—whether provided alone or alongside the problem statement—enhances their code generation capabilities.
    \item \textbf{Prompt Generic Tests (200/315):} This style, characterized by test cases with masked function names, posed the most challenges for the models, resulting in the lowest overall score. The reduced score in this category underscores the difficulty LLMs face when confronted with ambiguity. Without explicit function names or a clear problem statement, the models are tasked with deducing the problem's requirements and dealing with the added layer of uncertainty introduced by the masked function names. These tests were more intended to test the deductive reasoning ability of LLMs and less as a model for prompt engineers and LLM users to utilize.
\end{enumerate}

\subsection{Overall Performance by Model}

\begin{enumerate}
    \item \textbf{GPT-3.5} stands out as the top performer with a total score of 210/252, closely followed by \textbf{GPT-4} with 209/252. \textbf{Claude} is not far behind with a total of 204/252. This indicates that these models have a relatively similar and high level of adaptability and comprehension when tasked with code generation under varying prompting styles. This result could be attributed to the relative simplicity of the Python coding problems, as GPT-4 generally outperforms GPT-3.5 in most benchmarks \cite{unknown}.
    \item \textbf{Text-davinci-002} clearly lags behind the other models with a considerably lower score of 86/252, suggesting that it might either struggle with the coding tasks presented or require different prompting strategies to optimize its performance. Given that this model is the least advanced, this result was expected.
\end{enumerate}

\subsection{Analysis by Prompt Type}

\begin{enumerate}
\item \textbf{Prompt Generic Tests:} \textbf{GPT-4} performs the best in this category with 51/63, indicating its strong capability to adapt to ambiguity and infer requirements even when function names are masked. Given that GPT-4 is one of the state-of-the-art large language models capable of advanced reasoning, it was expected to emerge as the leader.
    \item \textbf{Prompt Only:} \textbf{Claude}, with a score of 54/63, showcases its proficiency in interpreting and generating code from straightforward, unembellished instructions.
    \item \textbf{Prompt Tests Only:} Both \textbf{GPT-3.5} and \textbf{GPT-4} share the top spot in this category, each scoring 55/63. This suggests that these models possess strong deductive reasoning skills, able to infer problem requirements solely based on test cases and their names.
    \item \textbf{Prompt with Tests:} \textbf{Claude}, \textbf{GPT-4}, and \textbf{GPT-3.5} all perform similarly well in this category with scores of 56/63 and 55/63 respectively, indicating their ability to effectively utilize additional contextual information in the form of test cases.

\end{enumerate}

\subsection{Comparing Models}

\begin{enumerate}
    \item \textbf{Bard} showcases consistent performance across all prompt types, with scores ranging from 36/63 to 46/63. Its performance indicates a balanced adaptability to different prompting styles and overall poor performance compared to other state-of-the-art models.
    \item \textbf{Claude} excels particularly when provided with more specific prompts such as "Prompt Only" and "Prompt with Tests," suggesting that it benefits from clearer and more detailed instructions.
    \item \textbf{GPT-4} and \textbf{GPT-3.5} display remarkable adaptability across all prompt styles, indicating their robustness and versatility in code generation tasks.
    \item \textbf{Text-davinci-002} demonstrates its age, as its performance is significantly below the other models across all prompting styles.
\end{enumerate}

\subsection{Implications}

The results imply that while all models have demonstrated the capability to generate code based on varying prompt styles, there are clear distinctions in their adaptability and proficiency. It suggests that while models like GPT-3.5 and GPT-4 are highly versatile and can handle a wide range of prompting styles effectively, models like Claude might benefit more from explicit and detailed instructions. Text-davinci-002's performance emphasizes the rapid development speed in LLMs and the previous importance placed on optimizing for either code or text-based tasks.

\section{Future Work}

The findings reported in this paper assess how varying levels of prompt specificity influence the performance of leading LLMs in generating Python code. While our research has provided valuable insights into these models' adaptability and comprehension skills, several avenues remain unexplored, offering potential for further research and deeper understanding. Here are some promising directions for future work:

\subsection{Exploration of Different LLMs}

While this study primarily focused on models like GPT-3.5, GPT-4, Claude-2, and Bard, the landscape of Large Language Models is vast and constantly evolving. Future research could explore the performance of other LLMs, including Llama 2 and Google’s yet-to-be-released Gemini, to provide a more comprehensive perspective on the capabilities and limitations of different models in automated code generation tasks.

\subsection{Testing in Different Programming Languages}

Our research predominantly concentrated on Python code generation. However, the versatility and adaptability of LLMs could be further tested by exploring code generation in different programming languages. This would elucidate whether the findings of our study are specific to Python or generalizable across various languages, offering insights into the universality of LLM capabilities. Evaluating the impact of prompt specificity on different language architectures, like functional languages, could also provide helpful insight into the reasoning ability of LLMs across programming paradigms. Moreover, looking at less popular languages could prove interesting, as the LLMs’ abilities to reason on generic tests may not exist in a language that the LLM has less training data about.

\subsection{Examination of More Complex Problems}

The coding problems employed in this study were diverse yet generally simple one-function programs. There is potential to delve into more intricate and complex problems in future investigations. By challenging LLMs with multifaceted problems that require deeper logic, advanced algorithms, or specialized knowledge, we could gauge the depth of understanding, logical reasoning, and problem-solving skills of these models.

\subsection{Experimentation with Varied Prompt Types}

Our study employed four distinct styles of prompts. Future research could experiment with different types of prompts, exploring varying degrees of specificity, ambiguity, or even introducing elements like visual cues, multi-modal inputs, or real-world context. This would allow for a more granular examination of how LLMs interpret and respond to diverse and unconventional prompting strategies.

\subsection{Exploring existing LLM Benchmarks}

There are numerous ways to benchmark a coding problem output, and several have been created specifically for LLMs. It would be beneficial to analyze Large Language Model output on various grading criteria, including style, space complexity, and time complexity.

\section{Concluding Remarks}
The following are the lessons we learned from conducting this research:
\begin{itemize}
    \item Prompting with tests enhances LLM code generation capabilities. LLMs can effectively leverage test cases to infer requirements and generate more accurate code, whether tests are provided alone or with problem statements. Consequently, software engineers utilizing AI-assisted tools for generating code would benefit from including test cases when prompting LLMs and adhering to a TDD approach throughout the development lifecycle.
    \item Ambiguity poses challenges for LLMs. Masked function names reduced performance, indicating difficulty adapting to uncertainty about specific requirements. While this was expected, it offers a benchmark by which future models can be evaluated on reasoning and deduction.
    \item Vast opportunities exist to build on this research. While our examination was limited to Python and constrained problem scopes, the insights showed existing LLMs' reasoning skills and limitations. There remains ample opportunity for future work to expand the breadth and depth of LLM evaluations through more complex problems, diverse languages, unconventional prompts, and emerging models. Another area to explore more deeply would be including different levels of context within the prompt types, such as other numbers of test cases, and following a more traditional chain-of-thought prompting approach modified for code generation. As LLMs evolve at a remarkable pace, prompt engineering will play an increasingly crucial role in effectively harnessing these models and realizing their full potential across a vast array of real-world applications.
    \item Of the models examined, GPT-3.5 and GPT-4 display remarkable proficiency across all prompt types, showcasing robust adaptability. Meanwhile, Claude performs well when provided with precise prompts, suggesting it benefits from more explicit instructions. The outdated Text-davinci-002 significantly lags behind all other models, confirming the rapid evolution of LLM architectures.
    \item Set realistic expectations when applying LLMs. Despite impressive advances, LLMs may still generate logically unsound code or fail on atypical edge cases. Judicious human oversight is critical, including testing, validation, analyzing generated code quality, and handling exceptions. Use LLMs to accelerate development, but employ proper code reviews and monitoring.

\end{itemize}

This study presented a comprehensive investigation into the performance of popular LLMs, such as Bard, Claude, GPT-3.5, and GPT-4, in generating Python code based on prompts with varying levels of specificity. By carefully designing a benchmark encompassing 104 coding problems and four distinct prompt types, we systematically evaluated the accuracy, efficiency, and reliability of the code generated by these models under different prompting conditions.

\section{Acknowledgments}
This paper used GPT-4 for assistance in generating content, code, and prompts, with original prompts created by Lincoln Murr.

\bibliographystyle{plain} 
\bibliography{sources} 

\section{Appendix}
The GitHub repository available at \url{github.com/murrlincoln/SWE-AI-Mini-Research} contains the test cases, prompts, and other relevant information presented in this paper. 

\end{document}